\documentclass[aps,prl,showpacs,twocolumn,amsmath,amssymb]{revtex4-1}
\usepackage{graphicx}
\usepackage{dcolumn}
\usepackage{bm}
\usepackage{amsmath}
\usepackage{amsfonts}
\usepackage{amssymb}
\usepackage{multirow}

\usepackage{amsthm}
\usepackage{pinlabel}
\usepackage{natbib}

\usepackage{epstopdf}
\usepackage[abs]{overpic}

\usepackage{cancel}

\usepackage[normalem]{ulem}

\usepackage[dvipsnames]{xcolor}

\usepackage{xr}

\usepackage{braket}

\definecolor{specialgray}{HTML}{505050}
\definecolor{col10K}{HTML}{FFA000}
\definecolor{col300K}{HTML}{924FA4}
\definecolor{colMu}{HTML}{5278BD}
\definecolor{colMuI}{HTML}{924FA4}
\definecolor{newred}{HTML}{D53E4F}
\definecolor{newblue}{HTML}{5278BD}
\definecolor{newcyan}{HTML}{4EBCB3}
\definecolor{newgreen}{HTML}{5CB14E}
\definecolor{newpurple}{HTML}{924FA4}
\definecolor{newyellow}{HTML}{D1C72E}
\definecolor{neworange}{HTML}{D6923C}

\usepackage{lineno}

\begin{document}
\title{Prominent Cooper Pairing Away From the Fermi Level 
 and its Spectroscopic Signature in Twisted Bilayer Graphene}
\author{Fabian Schrodi}
\author{Alex Aperis}\email{alex.aperis@physics.uu.se}
\author{Peter M. Oppeneer}
\affiliation{Department of Physics and Astronomy, Uppsala University, P.\ O.\ Box 516, SE-75120 Uppsala, Sweden}
	
\vskip 0.4cm
\date{\today}
	
\begin{abstract}
We investigate phonon-mediated Cooper pairing in flat electronic band systems by solving the full-bandwidth multiband Eliashberg equations for superconductivity in magic angle twisted bilayer graphene using a realistic tight-binding model. We find that Cooper pairing away from the Fermi level contributes decisively to superconductivity by enhancing the critical temperature and ensures a robust finite superfluid density. We show that this pairing yields particle-hole asymmetric superconducting domes in the temperature--gating phase diagram and gives rise to distinct spectroscopic signatures in the superconducting state. We predict several such features in tunneling and angle resolved photoemission  spectra for future experiments.
\end{abstract}
	
\maketitle

In the conventional theory of phonon-mediated superconductivity, as formulated by Eliashberg \cite{eliashberg1960}, Cooper pairing is expressed by the energy dependent function, $\Delta(\omega)$, which is proportional to the pair's condensation energy \cite{Scalapino1966}. When the electron energy scale dominates over the phonon one,  pairing takes place only within a narrow energy window, $\omega_c$, around the Fermi surface where $\Delta(\omega)>0$. If in addition the electron-phonon interaction (EPI) is very weak, one recovers the Bardeen-Cooper-Schrieffer (BCS) approximation, $\Delta(\omega)\simeq\Delta\times\theta(\omega_c-|\omega|)$ \cite{Carbotte1990}. 
Flat band systems \cite{Heikkilae2011} however present a fascinating counterexample to this picture. These materials exhibit extremely narrow electronic bands near the Fermi level that host Van Hove singularities in extended regions of reciprocal space. The flatness of these bands enables a reversed situation where the phonon energy scale dominates over the electron one \cite{Ikeda1992}. 
It is as yet unclear how the BCS picture applies to flat band superconductors. For example, a BCS approximation implies the peculiar absence of a robust Meissner effect unless extra geometrical terms contribute to the superfluid density \cite{Peotta2015}.

In this work, we unravel the specifics of Cooper pairing in flat band superconductors by providing a full solution to the appropriate Eliashberg equations \cite{Aperis2018,Gastiasoro2019}. We focus on magic angle twisted bilayer graphene (TBG), a prototypical flat band superconductor, that has recently attracted tremendous research interest due to the rich variety of physical phenomena that result from a vastly changed electronic structure  depending on the twist angle \cite{Cao2018_1,Cao2018_2,Yankowitz2019,Yeh2014,Li2009,Po2019,Jung2014}. 
Most importantly for our present work, two nearly flat bands develop in close vicinity of the Fermi level \cite{Morell2010,Cao2018_1,Cao2018_2}. 
When gated, TBG becomes superconducting at the magic twist angle $\sim1.1^{\circ}$ with maximum critical temperature $T_c$ ranging between $0.5\,$K and $1.7\,$K  \cite{Cao2018_2,Yankowitz2019}. 
 There is a competition with external gating  between the superconducting and an insulating state \cite{Cao2018_2,Yankowitz2019}. The latter is associated with  enhanced correlations that   manifest as extended features seen by tunneling experiments in the non-superconducting state of TBG \cite{Xie2019a,Kerelsky2019,Choi2019}. Several works have recently emphasized  the relevance of the EPI for superconductivity in TBG   \cite{Wu2018,Peltonen2018,Choi2018,Lian2019,Ojajaervi2018}. However, to our knowledge, all previous theories focused on  effective descriptions near the Fermi level.

Here we present multiband, full-bandwidth Eliashberg calculations of superconductivity in TBG having as input 
realistic electron dispersions and EPI.
 We find that, in contrast to the BCS picture, $\Delta(\omega)>0$ throughout the full bandwidth and show that Cooper pairing stems primarily from electrons away from the Fermi level. By calculating the full temperature--doping superconducting phase diagram, we show that this type of pairing not only contributes to $T_c$, but also does this in a particle-hole asymmetric way that agrees with experimental observations  \cite{Cao2018_1,Yankowitz2019}. Furthermore, our calculations demonstrate that in flat band superconductors, Cooper pairing away from the Fermi level ensures a finite superfluid density. Finally, we calculate angle resolved photoemission spectroscopy (ARPES) and scanning tunneling spectroscopy (STS) spectra and predict distinct low energy features below $T_c$ as signatures of such pairing.

\begin{figure}[t!]
\includegraphics[width=\linewidth]{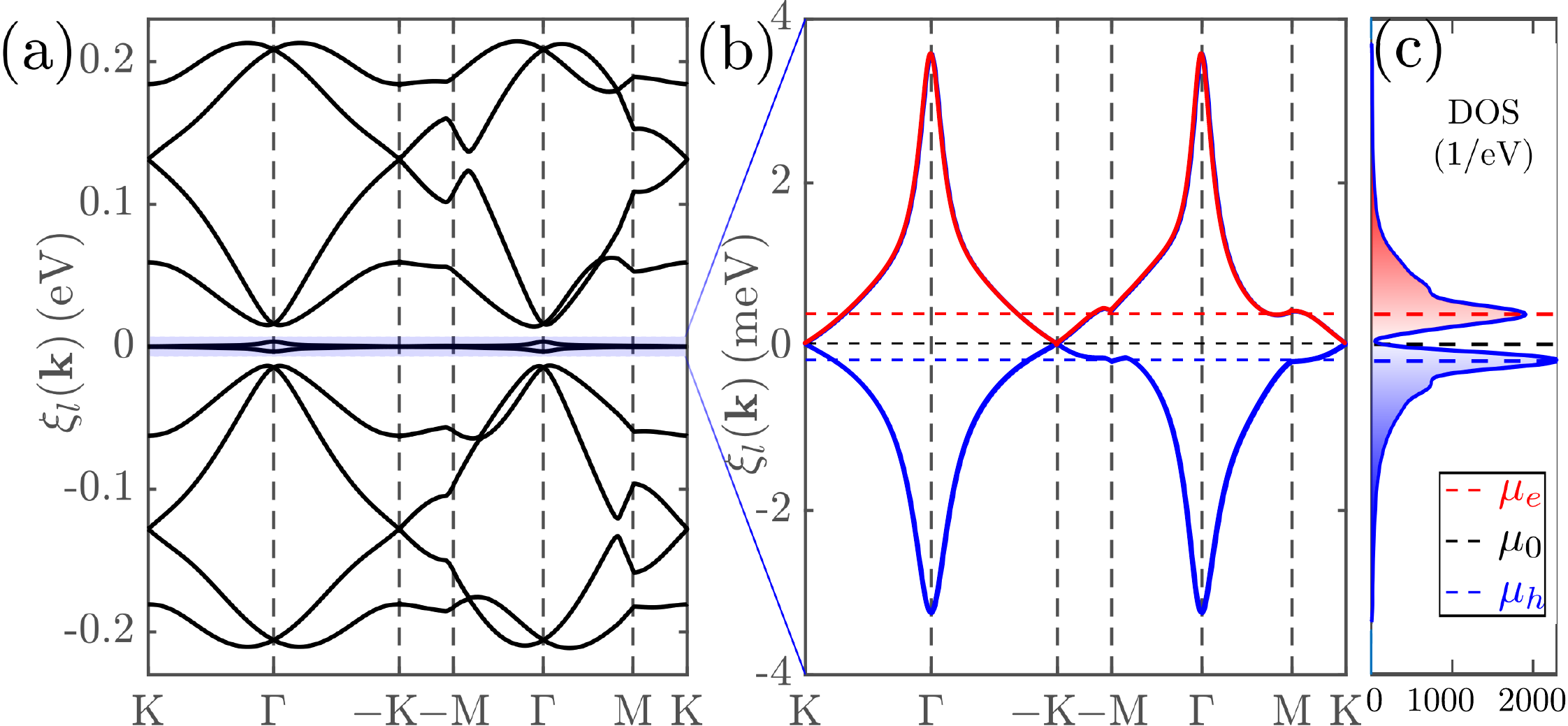}
\caption{(a) Ten band electronic dispersion for undoped ($\mu_0=0$) TBG at the magic angle $1.05^{\circ}$ \cite{Po2019}. (b) Zoom-in near the Fermi level (marked by a black dashed line) showing the two flat bands and (c) their corresponding DOS. The chemical potential needed to bring each flat band region and the corresponding peak in the DOS at the Fermi level is marked with red (blue) dashed lines for electron (hole) doping.}
\label{fig1}
\end{figure}

To accurately describe the electronic properties of magic angle TBG, we adopt a ten band tight-binding model \cite{Po2019} that provides a faithful band structure, $\xi_l(\mathbf{k})$, that is consistent with findings of the continuum model \cite{dosSantos2007,dosSantos2012,Bistritzer2011,Nam2017} and {\em ab initio} calculations \cite{Fang2016,Uchida2014,Jung2014}. Here $l$ is the band index and $\mathbf{k}$ the momentum in the mini Brillouin zone (BZ). A perturbation term in the Hamiltonian breaks the particle-hole symmetry and leads to a pair of nearly flat bands very close to the Fermi level, see Fig.\,\ref{fig1}(a). 
 These two bands are energetically well separated from the rest by over 20 meV and have a very narrow bandwidth, $W\approx7$ meV, so that they harbor flat regions with extremely high DOS near the M--K and (-M)--(-K) symmetry lines [Fig.\ref{fig1}(b)-(c)].
 As a result, the rest of the bands are marginally relevant and we hence confine ourselves here to these two flat bands in all following calculations. The inclusion of the complete ten bands has been carefully checked to not alter our results. As shown in Fig.\,\ref{fig1}(b), the two flat bands form Dirac points at the $\pm$K points of the BZ and exhibit Dirac-like cones at their band extrema, located at $\Gamma$. 

Whereas focus has been put on the role of EPI involving intralayer modes \cite{Wu2018,Peltonen2018,Choi2018,Lian2019}, significant interlayer coupling has also been calculated \cite{Choi2018}. 
Driven by recent experiments where superconductivity in TBG has been found to be tunable with the distance between the two graphene sheets \cite{Yankowitz2019}, here we assume as the mediator of superconductivity the interlayer breathing mode with the characteristic phonon frequency of $\Omega=11\,$meV\,\cite{Choi2018,Cocemasov2013}.
This mode can provide an attractive interaction between electrons in an energy window that is $\sim1.6\times$W.
Given this fact and the similarity between the two dispersions, we expect that electrons from these bands couple to the mode similarly, regardless of their band index. Thus, we assume a global electron-phonon coupling strength, $g_0$, which we also take as isotropic. By computing  $T_c$ and requesting it to match to experimental values \cite{Cao2018_2,Yankowitz2019} our Eliashberg calculations yield $g_0=1.5\,$meV, which is kept fixed henceforth. In what follows, we do not include Coulomb pair breaking effects explicitly but assume that these are incorporated in the value of $g_0$. Thus $g_0$ represents the effective attractive interaction strength.

To study our electron-phonon coupled system we employ full-bandwidth, multiband Eliashberg theory \cite{Aperis2018, Schrodi2018}. This theory takes explicitly into account scattering processes involving electrons with energies and momenta that are not restricted to the vicinity of the Fermi surface, therefore it goes beyond Migdal's theorem. 
 As such, it also allows the possibility of Cooper pairing away from the Fermi level to be taken fully into account \cite{Aperis2018,Gastiasoro2019}.
  The matrix self-energy 
using Pauli matrices $\hat{\rho}_i$, is
\begin{eqnarray}
\hat{\Sigma}(i\omega_m) =&& T\sum_{\mathbf{k},m'}\sum_{l} \frac{2g_0^2\Omega \cdot \hat{\rho}_3\hat{G}_{l}(\mathbf{k},i\omega_{m'})\hat{\rho}_3}{(\omega_m-\omega_{m'})^2+\Omega^2} ,~~\label{selfenergy}
\end{eqnarray}
with Matsubara frequencies $\omega_m$=$\pi T(2m+1)$, temperature $T$ and matrix Green's function,
\begin{eqnarray}
~~&\hat{G}^{-1}_l(\mathbf{k},i\omega_m) = \big[i\omega_mZ(i\omega_m) \hat{\rho}_0  - \phi(i\omega_m) \hat{\rho}_1 ~~~~~~~~~~~~~~~~ \nonumber\\
&- (\xi_l(\mathbf{k})-\mu+\chi(i\omega_m))\hat{\rho}_3 \big]\,.\label{greensfunction}
\end{eqnarray}
The chemical potential $\mu$ simulates gating effects. From the Dyson equation, we obtain a system of three coupled selfconsistent Eliashberg equations for the mass renormalization $Z(i\omega_m)$, pairing function $\phi(i\omega_m)$ and chemical potential renormalization $\chi(i\omega_m)$ that are complemented by an equation for the particle filling, $n$ (see Supplemental Material (SM)). The superconducting gap function has s-wave, spin singlet symmetry and is given by $\Delta(i\omega_m)=\phi(i\omega_m)/Z(i\omega_m)$. 
The symmetry of the superconducting order parameter in TBG is yet unresolved experimentally. Previous BCS theory calculations obtained both s-wave and d-wave \cite{Wu2018} or solely s-wave \cite{Lian2019} phonon-mediated superconductivity. Atomistic calculations \cite{Choi2018} and symmetry analysis studies based on available experimental data \cite{Talantsev2020} also support the s-wave picture while the latter study excludes d-wave symmetry.

We first solve selfconsistently the multiband full-bandwidth Eliashberg equations for different values of $(T,\mu)$ and subsequently map the results to the corresponding sets of $(T,n)$ values. The obtained Matsubara space results can be used to calculate any thermodynamic property that stems from the system's free energy. Here, this is done for the superfluid weight as will be discussed below. All quantities are then analytically continued from Matsubara to real frequencies by a selfconsistent procedure that is formally exact \cite{Marsiglio1988}. This gives us access to the precise real frequency dependent retarded Green's function at all temperatures and therefore to spectroscopic quantities of interest like the ones measured by STS and ARPES (see SM and Refs.\,\cite{Aperis2018,Schrodi2018}). All calculations are performed with the Uppsala Superconductivity (UppSC) code \cite{UppSC}.

\begin{figure}[t!]
	\centering
   \includegraphics[width=\linewidth]{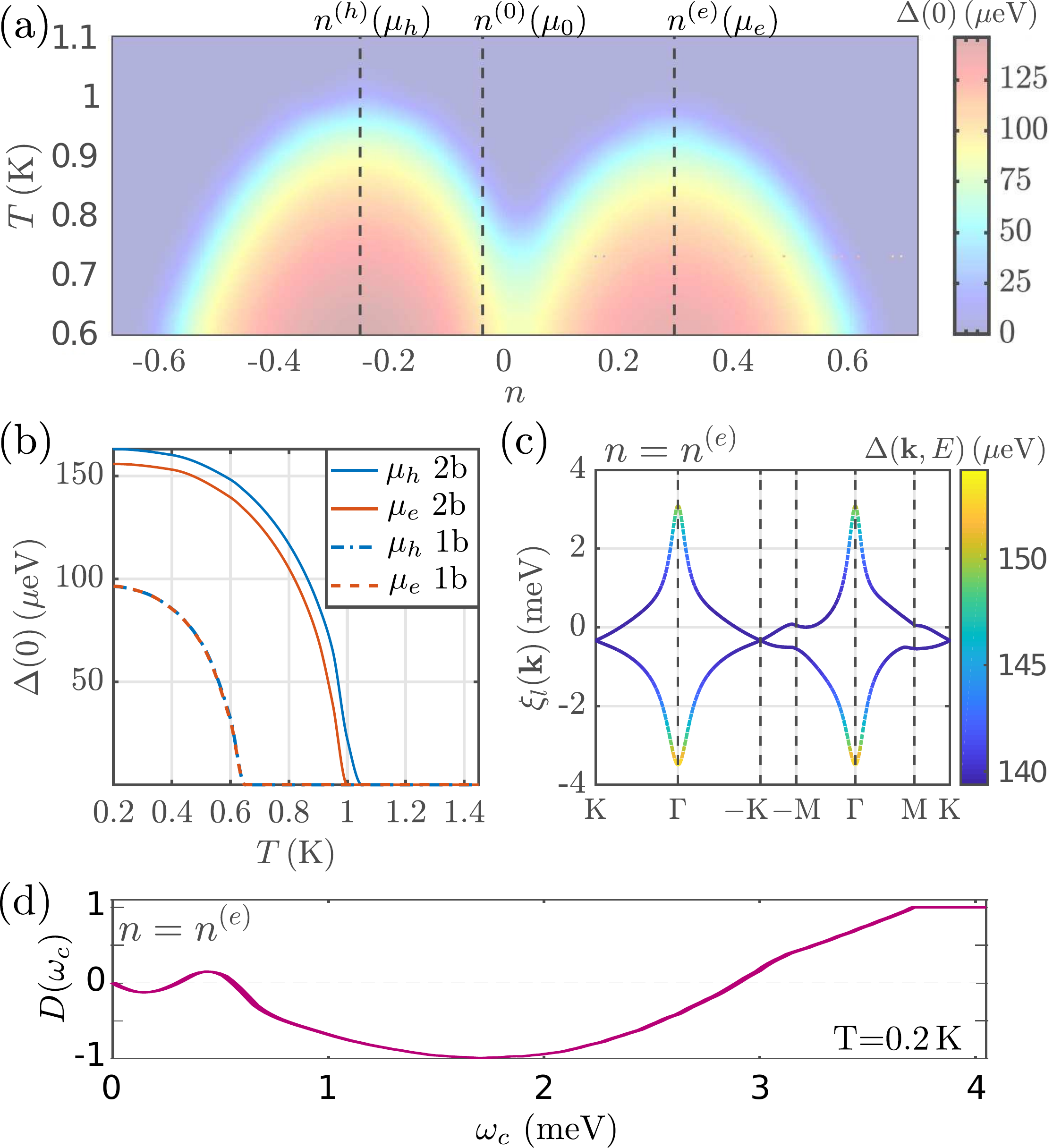}
	\caption{(a) Temperature--doping phase diagram of the zero-frequency superconducting gap, $\Delta(0)$. Highlighted are fillings of the two dome-centers and the value corresponding to $\mu=0$. (b) Temperature dependence of $\Delta(0)$ for optimal 
doping as indicated in the legend. `1b' (`2b') refer to one (two)-band calculations.
 (c) Superconducting gap projected on the renormalized band structure for $T=0.6\,$K and filling $n=n^{(e)}$. (d) Calculated normalized superfluid weight $D (\omega_c )$ for $n=n^{(e)}$ as a function of the cutoff, $\omega_c$.}
 \label{fig2}
\end{figure}

Fig.\,\ref{fig2}(a) shows our calculated temperature--filling phase diagram with respect to $\Delta(0)$. Here, $n^{(0)}$ is the electron filling of the system without gating/doping ($\mu=0$) and $n^{(h)}$ ($n^{(e)}$) are the reference points where the lower (upper) flat band is half-filled and the corresponding DOS peaks (see Fig.\ref{fig1}(b)-(c)), which happens for hole or electron doping, i.e.\ $\mu_h<0$ or $\mu_e>0$. For hole doping ($n<0$) we resolve a superconducting dome in qualitative agreement with experiment \cite{Cao2018_1,Yankowitz2019}. The counterpart for electron doping ($n>0$) exhibits a lower $T_c$ as has also been measured \cite{Yankowitz2019}. Near half-filling ($n^{(0)}$) where insulating states are observed \cite{Cao2016,Cao2018_1,Yankowitz2019,Jiang2019}, we find $T_c$ $\neq$ 0. This is not surprising since particle-hole pairing is not included in our model. Here we focus on the superconducting domes that have been observed experimentally and are well described by our theory.

With our chosen parameter set we find a maximum gap of $\Delta\sim163\,\mu$eV and $T_c\sim1.05\,$K for for $n^{(h)}$, whereas for $n^{(e)}$, $\Delta\sim154\,\mu$eV and $T_c\sim1\,$K. Clearly, $T_c$ is maximized when the doping level is such that the flat portion of the bands lies on the Fermi level, i.e.\ when the Fermi level DOS is maximized. This may  be expected given that the coupling strength is proportional to the DOS  \cite{Wu2018,Peltonen2018}. A similar argument could also be conjectured to explain the electron-hole asymmetry in the shape of the two domes of Fig.\,\ref{fig2}(a), since the particle-hole asymmetry of the TBG bandstructure leads to DOS peaks that differ significantly in height, see Fig.\ref{fig1}(c). Notably, such electron-hole asymmetry was not found in previous BCS calculations of the TBG phase diagram, despite the fact that the input bandstructure was particle-hole asymmetric \cite{Wu2018,Peltonen2018}.

To shed more light on this apparent discrepancy, we repeated our Eliashberg calculations for the same set of parameters but now taking as input only one of the flat bands at a time. In these single-band calculations we first keep only the upper band (red line in Fig.\ref{fig1}(b)) and place the Fermi level at $\mu_e$ and subsequently take only the lower band (blue line in Fig.\ref{fig1}(b)) and place the Fermi level at $\mu_h$. The obtained results are shown in Fig.\,\ref{fig2}(b) with dashed (solid) lines for one (two) band calculations.
Remarkably, the one band results are identical to each other despite the particle-hole asymmetry between the bands and the pronounced difference in Fermi level DOS. The superconducting gap and the critical temperature are reduced by approximately $40\%$ as compared to the full calculation. It is only when we consider both bands together that  $T_c$ is enhanced in a particle-hole asymmetric way. These results attest that at a given doping, the band not crossing the Fermi level is always involved in superconductivity, which points to the relevance of Cooper pairing away from the Fermi level in TBG. Such type of pairing has been recently predicted in FeSe/SrTiO$_3$ \cite{Aperis2018} where the phonon energy scale is comparable to the electron one. 

Having calculated the precise real frequency dependence of all the quantities involved in our theory, we proceed 
by projecting the superconducting gap function on the underlying renormalized band structure, $\tilde{\xi}_l({\bf k})$ \cite{Aperis2018}. 
The latter is a solution to $\tilde{\xi} = [\xi_l(\mathbf{k})-\mu+\chi(\tilde{\xi})]/Z(\tilde{\xi})$ for each $\mathbf{k}$ and $l$. We then map $\Delta(\omega)$ to $\Delta(\tilde{\xi}_l({\bf k}))$.
Fig.\,\ref{fig2}(c) shows this projection for $n=n^{(e)}$ along high-symmetry lines. Remarkably, the maximum $\Delta$ value is located 
at the largest energy away from the Fermi level. This behavior is  markedly different from the conventional picture where $\Delta<0$ above a given energy, due to the EPI becoming anti-pairing for electrons at such high energies \cite{Scalapino1966}. In our case, $\Delta$ stays positive (i.e.\ the EPI is pairing) \textit{throughout the full bandwidth}. Thus, our results show that the Cooper pairing in TBG stems prevalently from electrons away from the Fermi level. These Cooper pairs have even frequency, s-wave, spin singlet symmetry. Both bands, irrespective of which one crosses the Fermi level, contribute significantly to the Cooper pair formation and hence to $T_c$.

Given the above results, it is now instructive to calculate the so-called London kernel, 
\begin{align}
&Q_{\alpha\beta}(T) = e^2 T \sum_{\mathbf{k},l,m} \left( \big(\nabla^2_{\alpha\beta}\xi_l(\mathbf{k})\big) \mathrm{Tr}\big[\hat{\rho}_3\hat{G}_l(\mathbf{k},i\omega_m)\big] \right. \nonumber\\\label{london}
&\left. + \big(\nabla_{\alpha}\xi_l(\mathbf{k})\big)\big(\nabla_{\beta}\xi_l(\mathbf{k})\big)\mathrm{Tr}\big[\hat{G}_l(\mathbf{k},i\omega_m)\hat{G}_l(\mathbf{k},i\omega_m)\big]\right),
\end{align}
that describes the local, static current response to an applied transverse vector potential, $J_\alpha=-Q_{\alpha\beta}A^\beta$, with $\alpha,\beta=x,y$. This is proportional to the superfluid density, $n_S$ (or superfluid weight), which in turn is related to the penetration depth via $n_S\propto\lambda^{-2}$. The first (second) summand in Eq.\,(\ref{london}) is the diamagnetic (paramagnetic) part. Taking the usual approximations, Eq.\,(\ref{london}) reduces to the standard energy-integrated isotropic Eliashberg form \cite{Marsiglio1990, Golubov2002}.
 However, here we solve Eq.\,(\ref{london}) as is. We also note that no extra  geometrical (topological) term \cite{Peotta2015} is included here.

The diamagnetic term is usually considered vanishingly small, so that one often focuses on the paramagnetic part which is proportional to $\Delta^2$. For a flat band system, with an electronic dispersion of the general form $\xi_{\bf k}\propto \pm({\bf k}/{\bf k}_{fb})^N$, where $N \gg 2$ and ${\bf k}_{fb}$ controls the extent of the flat band region \cite{Heikkilae2011}, Eq.\,(\ref{london}) would yield $Q\approx 0$ if one naively assumes that $\Delta\rightarrow 0$ for $|\xi_{\bf k}|>\xi_{{\bf k}_{fb}}$. This is because the electron velocity is vanishingly small for $|\xi_{\bf k}|<\xi_{{\bf k}_{fb}}$ where $\Delta$ would be expected to be large. Such a situation would appear problematic since then $n_S\rightarrow 0$ and $\lambda\rightarrow\infty$ and it has been proposed that 
topological terms beyond Eq.\,(\ref{london}) would become essential to produce $n_s\neq 0$ \cite{Peotta2015}.
On the other hand, for momenta away from the flat band region and therefore away from the Fermi level, the rapid increase in velocities would lead to a significant contribution in Eq.\,(\ref{london}) if $\Delta$ remains large for $|\xi_{\bf k}|>\xi_{{\bf k}_{fb}}$. This is exactly what we find here, as shown in Fig.\ref{fig2}(b),(c). To quantify this observation, we introduce the normalized superfluid weight $D(\omega_c)=\sum_{\bf k}^{|\xi_{\bf k}|\leq\omega_c}Q_{xx}({\bf k})/Q_{xx}$, where $Q_{xx}({\bf k})$ is the ${\bf k}$-summand of Eq.\,(\ref{london}) and $\omega_c$ is an energy cutoff. As shown in Fig.\ref{fig2}(d), for small-$\omega_c$, where processes only at the Fermi surface enter into Eq.\,(\ref{london}), $D(\omega_c)\approx 0$. However, for $\omega_c\geq\rm{max}|\xi_l({\bf k})|$, we retrieve the full calculation of Eq.\,(\ref{london}) so that $D(\omega_c)=1$. Thus, there is no obstruction for having a finite superfluid density or a well defined Meissner effect in TBG \cite{Xie2019,Julku2020,Hu2019} or in any similar flat band system that deviates from the strictly flat limit \cite{Peotta2015}. Our Eliashberg theory-predicted Cooper pairing lends support to recent model calculations where the non-geometric contribution to the superfluid density is found to dominate at a given parameter range \cite{Julku2020}.

\begin{figure}[t!]
	\centering
	\includegraphics[width=1\columnwidth]{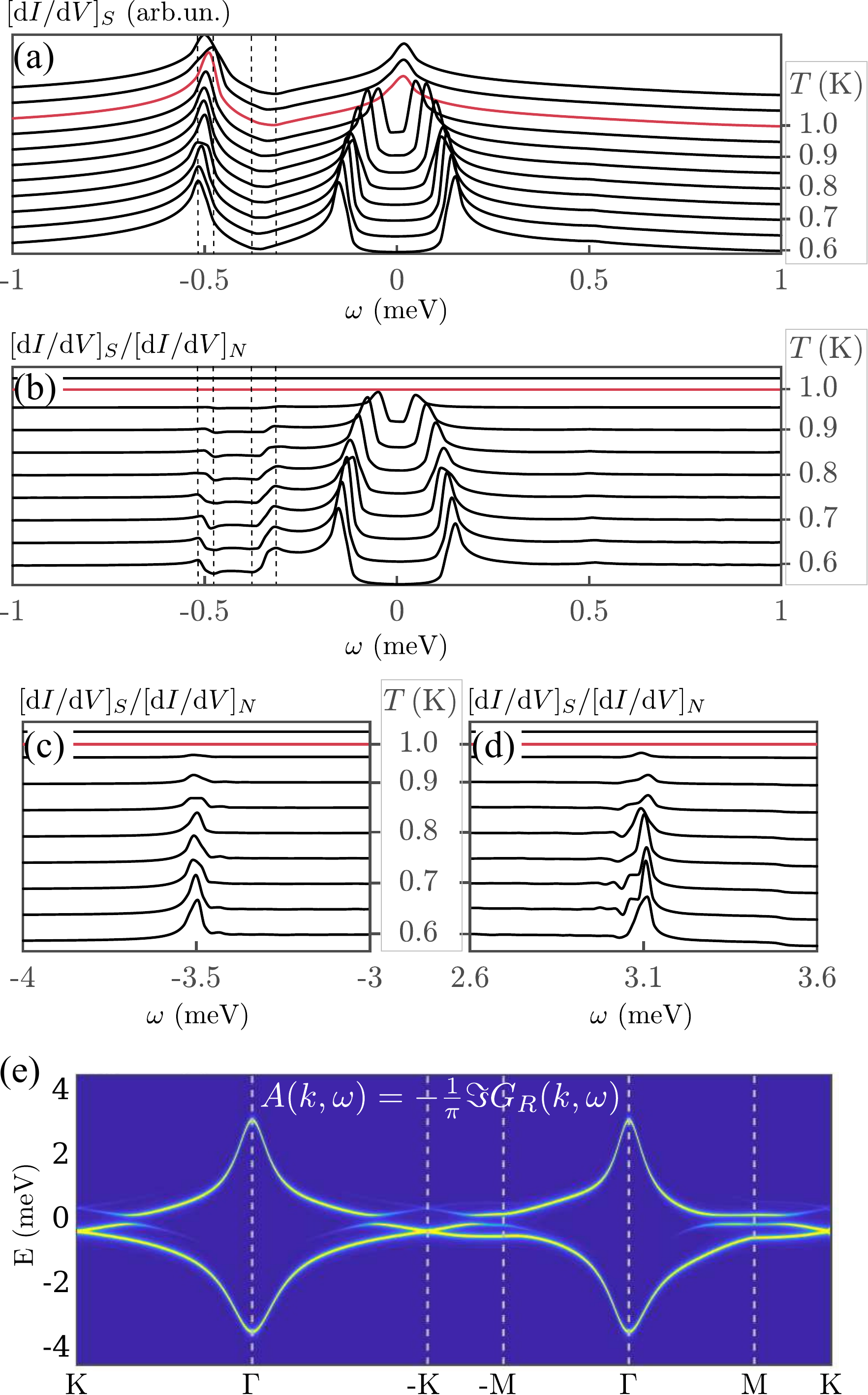}
	\caption{Calculated spectra for filling $n=n^{(e)}$ and different temperatures. Here $T_c=1.0\,\mathrm{K}$. (a) Non-normalized tunneling and (b) normalized tunneling spectra for  $T\in[0.6,1.05]$. (c) and (d) Normalized tunneling for energies near the band edges of the flat bands. (e) Spectral function along high-symmetry lines of the BZ at T=0.2\,K, relevant to ARPES.}\label{fig4}
\end{figure}

We now turn to the experimental signatures of our predicted Cooper pairing in TBG by using our obtained real frequency Green function to calculate STS and ARPES spectra (see SM). So far, no ARPES measurements exist for TBG and the existing STS studies have mainly focused on the non-superconducting spectral properties \cite{Luican2011,Yan2012,Wong2015,Li2009,Xie2019a,
Kerelsky2019,Choi2019}. Here we present results for $n=n^{(e)}$. Results for $n=n^{(h)}$ are similar and included in the SM. Fig.\ref{fig4}(a) shows the differential conductance, ${\rm d}I/{\rm d}V\propto\rm{DOS}$, for several temperatures below and above $T_c$. At $T>T_c$, the shape of the DOS is modified compared to Fig.\ref{fig1}(c) due to the included EPI, but the relative height of the peaks remains similar. Below $T_c$, the sharp peak at the Fermi level gives way to two coherence peaks separated by a gap, characteristic of s-wave superconductivity. The peak at negative bias that corresponds to the flat band region of the band below the Fermi level remains sharp. However, closer inspection reveals that it slightly moves to the left as temperature decreases. When we now consider the quasiparticle energy spectrum in the superconducting state,
\begin{eqnarray}\label{poles}
\omega=\pm\Re\sqrt{ \left(\xi_l(\mathbf{k})-\mu+\chi(\omega)\right)^2/Z^2(\omega) + \Delta^2(\omega) } \, ,
\end{eqnarray}
it becomes evident that this peak shift is a manifestation of having a non-decreasing $\Delta$ away from the Fermi level.

Fig.\ref{fig4}(b) shows the renormalized tunneling spectrum that is the ratio of the spectra below, $[{\rm d}I/{\rm d}V]_S$, and above $T_c$, $[{\rm d}I/ {\rm d}V]_N$. Such experimentally accessible spectra are ideal for providing finer details of the modified tunneling due to superconductivity and therefore for identifying the  signatures of our predicted Cooper pairing away from the Fermi level. Indeed, Fig.\ref{fig4}(b) shows that apart from the coherence peaks around zero bias, there appears a hump-dip and a dip-hump for $\omega \ll 0$. Clearly, these structures occur as a result of the relative displacement between $[{\rm d}I/{\rm d}V]_S$ and $[ {\rm d}I/ {\rm d}V]_N$ due to Cooper pairing, see e.g.\ Eq.\,(\ref{poles}). Similar arguments apply when we look at frequency regions corresponding to the upper and lower edges of the electronic dispersion, shown in Fig.\,\ref{fig4}(c)-(d)\,. At the band edge, the DOS is minimal whereas $\Delta$ acquires its maximum value. Therefore, the renormalized spectra exhibit sharp peaks there that should be experimentally discernible. We also find a less pronounced dip-hump near $+0.5\,$meV [see Fig.\ref{fig4}(b)] which could be experimentally detectable via a second-derivative analysis. This feature appears as the particle-hole symmetric replica of the structure near $\omega=-0.5\,$meV and is due to the twofold Bogoliubov quasiparticle energy dispersion in the superconducting state [cf.\,Eq.\,(\ref{poles})]. 

The quasiparticle spectrum can be observed with ARPES that probes directly the momentum and energy resolved spectral function, $A({\bf k},\omega)$. Our calculated $A({\bf k},\omega)$ is shown in Fig.\ref{fig4}(e) for $n=n^{(e)}$ at ${T} \ll {T}_c$. In Fig.\,\ref{fig4}(e), the Bogoliubov bands are clearly resolved to exhibit the characteristic ``back-bending'' near the Fermi level. Signatures of $\Delta$ remaining large away from the Fermi level are not easy to discern here. However, we observe that the Dirac crossing at K (-K) has a Bogoliubov replica. In the case of hole doped TBG, this replicated Dirac crossing lies below the Fermi level (see SM) and, with sufficient resolution, can in principle be measured. 

In conclusion, our full-bandwidth multiband Eliashberg calculations for phonon-mediated superconductivity in magic angle twisted bilayer graphene reveal that the Cooper pairing develops primarily away from the Fermi level and contributes significantly to the $T_c$ \cite{Aperis2018}. The existence of such Cooper pairs provides an alternative solution to the vanishing superfluid density paradox that was first encountered in strictly flat-band superconductors \cite{Peotta2015}.
 In addition, they give rise to nontrivial spectroscopic features at low energy and temperature that should be distinct from the ones already observed in the non-superconducting state \cite{Xie2019a,Kerelsky2019,Choi2019}. These are summarized as follows: new features in the normalized  tunneling that occur a) away from the superconducting  coherence peaks and near the location of flat-band regions, b) far away from the superconducting coherence peaks and at the location of the electron band edges, and c) replica bands in the ARPES spectra that occur due to extended energy region of the Bogoliubov bands in the superconducting state. Our predictions for the superconducting spectra provide a route for the unambiguous identification of Cooper pairing away from the Fermi level. Our findings should be qualitatively applicable for all phonon-mediated flat band superconductors \cite{Heikkilae2011,Kopnin2011,Wang2017}.

\begin{acknowledgments}
We are grateful to P\"aivi T\"orm\"a for valuable discussions. We acknowledge support from the Swedish Research Council (VR), the R{\"o}ntgen-{\AA}ngstr{\"o}m Cluster, the K.\ and A.\ Wallenberg Foundation (Grant No.\ 2015.0060) and the Swedish National Infrastructure for Computing (SNIC). 
\end{acknowledgments}

\bibliographystyle{apsrev4-1}

\appendix

\newpage

\begin{widetext}
\newpage
\section{\large Supplemental Material for ``Prominent Cooper Pairing Away From the Fermi Level and its Spectroscopic Signature 
in Twisted Bilayer Graphene''}
\author{Fabian Schrodi}
\author{Alex Aperis}\email{alex.aperis@physics.uu.se}
\author{Peter M. Oppeneer}
\affiliation{Department of Physics and Astronomy, Uppsala University, P.\ O.\ Box 516, SE-75120 Uppsala, Sweden}
\begin{center}
Fabian Schrodi, Alex Aperis$^*$, and Peter M. Oppeneer\\
\textit{Department of Physics and Astronomy, Uppsala University, P.\ O.\ Box 516, SE-75120 Uppsala, Sweden}\\
(Dated: \today)
\end{center}
\end{widetext}

\section{Eliashberg theory}

In this section we give a short derivation of the Eliashberg equations that are solved in this work. The microscopic Hamiltonian of the system can be expressed as 
\begin{align}
H =& \sum_{\mathbf{k},l}\xi_{l}(\mathbf{k})\Psi^{\dagger}_{\mathbf{k},l} \hat{\rho}_3\Psi_{\mathbf{k},l} + \sum_{\mathbf{q}}\hbar\Omega \left(b^{\dagger}_{\mathbf{q}}b_{\mathbf{q}} + \frac{1}{2}\right) \nonumber \\
&+ \sum_{\mathbf{k},\mathbf{k}'}\sum_{l,l'}g_{\mathbf{q}}u_{\mathbf{q}}\Psi^{\dagger}_{\mathbf{k}',l}\hat{\rho}_3\Psi_{\mathbf{k},l'} ~.
\end{align}
It consists of an electron and a phonon part, as well as a term for the electron-phonon coupling. $\mathbf{k}$ and $\mathbf{q}$ denote wave vectors, $\xi_l(\mathbf{k})$ are electronic energies, $\hat{\rho}_i$ are Pauli matrices and $u_{\mathbf{q}}$ are atomic displacements. Electrons in band $l$ with spin $\sigma$, and phonons are  created (annihilated), respectively, by $c_{\mathbf{k},l,\sigma}^{\dagger}$ ($c_{\mathbf{k},l,\sigma}$) and $b_{\mathbf{q}}^{\dagger}$ ($b_{\mathbf{q}}$). The spin index is hidden in Nambu spinors $\Psi^{\dagger}_{\mathbf{k},l} = \big(c_{\mathbf{k},l,\uparrow}^{\dagger},c_{-\mathbf{k},l,\downarrow}\big)$. We can write the electronic self-energy as
\begin{eqnarray}
\hat{\Sigma}(i\omega_m) =&& T\sum_{\mathbf{k},m'}\sum_{l'}\hat{\rho}_3\hat{G}_{l'}(\mathbf{k},i\omega_{m'})\hat{\rho}_3 \nonumber\\
&&\times \int_0^{\infty}\mathrm{d}\omega \frac{\alpha^2F_{l'}(\omega)}{N_{l'}(0)}\frac{2\omega}{(\omega_m-\omega_{m'})^2+\omega^2} ,~~\label{selfenergy}
\end{eqnarray}
which corresponds to the `rainbow' Feynman diagram for electron-phonon scattering. At temperature $T$ the fermionic Matsubara frequencies are given by $\omega_m=\pi T(2m+1)$, $m\in\mathbb{Z}$. For the Einstein frequency spectrum that is assumed here, the Eliashberg function can be written as
\begin{eqnarray}
\alpha^2F_l(\omega)=N_l(0) |g_0|^2\delta(\omega-\Omega) \, ,
\end{eqnarray}
with $N_l(0)$ the density of states at the Fermi level. 
We treat the scattering matrix elements as isotropic, i.e.\ $|g_{\mathbf{q}}|^2=|g_0|^2$.

Introducing a chemical potential $\mu$ that rigidly shifts the electronic energies, the Green's function in Nambu space is given by
\begin{eqnarray}
~~&\hat{G}_l(\mathbf{k},i\omega_m) = \big[i\omega_mZ(i\omega_m) \hat{\rho}_0  - \phi(i\omega_m) \hat{\rho}_1 ~~~~~~~~~~~~~~~~ \nonumber\\
&- (\xi_l(\mathbf{k})-\mu+\chi(i\omega_m)))\hat{\rho}_3 \big] \Theta^{-1}_l(\mathbf{k},i\omega_m) ~, \label{greensfunction}
\end{eqnarray}
with
\begin{eqnarray}
&\Theta_l(\mathbf{k},i\omega_m) = -\omega_m^2Z^2(i\omega_m) - \phi^2(i\omega_m)  ~~~~~~~~~~~~~~~~ \nonumber\\
&- (\xi_l(\mathbf{k})-\mu+\chi(i\omega_m))^2 ~,
\end{eqnarray}
where the mass renormalization $Z(i\omega_m)$, the gap function $\phi(i\omega_m)$, and chemical potential renormalization $\chi(i\omega_m)$ are momentum independent due to the underlying isotropy of the electron-phonon interaction. Next we use the self-energy of Eq.\,(\ref{selfenergy}), together with Eq.\,(\ref{greensfunction}), in the Dyson-Gorkov equation $\big[\hat{G}_l(\mathbf{k},i\omega_m)\big]^{-1}=\big[\hat{G}^0_l(\mathbf{k},i\omega_m)\big]^{-1}-\hat{\Sigma}(i\omega_m)$, where the Green's function of the non-interacting system is defined via
\begin{eqnarray}
\left[\hat{G}^0_l(\mathbf{k},i\omega_m)\right]^{-1} = i\omega_m\hat{\rho}_0 - (\xi_l(\mathbf{k})-\mu)\hat{\rho}_3 ~.
\end{eqnarray}
This leads to a set of coupled selfconsistent Eliashberg equations
\begin{eqnarray}
Z(i\omega_m) &=& 1-\frac{T}{\omega_m} \sum_{\mathbf{k},m',l} V^{\mathrm{e-ph}}(q_{m-m'})\frac{\omega_{m'}Z(i\omega_{m'})}{\Theta_l(\mathbf{k},i\omega_{m'})} ,~~ \label{z} \\
\chi(i\omega_m) &=& T \sum_{\mathbf{k},m',l} V^{\mathrm{e-ph}}(q_{m-m'}) \nonumber\\
&&~~~~~~~~~\times\frac{\xi_l(\mathbf{k})-\mu+\chi(i\omega_{m'})}{\Theta_l(\mathbf{k},i\omega_{m'})} ,~\\
\phi(i\omega_m) &=& -T \sum_{\mathbf{k},m',l} V^{\mathrm{e-ph}}(q_{m-m'})\frac{\phi(i\omega_{m'})}{\Theta_l(\mathbf{k},i\omega_{m'})} ~, \label{phi}
\end{eqnarray}
where we use the interaction kernel
\begin{eqnarray}
V^{\mathrm{e-ph}}(q_{m-m'}) = \int_0^{\infty}\mathrm{d}\omega \frac{\alpha^2F_{l'}(\omega)}{N_{l'}(0)}\frac{2\omega}{q_{m-m'}^2+\omega^2} ~,~~
\end{eqnarray}
with bosonic frequencies $q_m=2\pi Tm$. The set of Eliashberg equations (\ref{z}-\ref{phi}) is complemented by an equation for the electronic band filling
\begin{eqnarray}
n = 1 - \frac{2T}{L}\sum_{\mathbf{k},m}\sum_l \frac{\xi_l(\mathbf{k})-\mu+\chi(i\omega_m)}{\Theta_l(\mathbf{k},i\omega_m)} \, ,\label{n1}
\end{eqnarray}
where $L$ denotes the number of electronic bands.

Once we have solved selfconsistently for the Matsubara space functions $Z(i\omega_m)$, $\chi(i\omega_m)$, and $\phi(i\omega_m)$, we can calculate ARPES and STS spectra \cite{Aperis2018} to make direct contact with experiment. For this purpose we analytically continue the solutions to Eqs.\,(\ref{z}-\ref{phi}) selfconsistently via
\begin{eqnarray}
&&Z(\omega) = 1 - \frac{T}{\omega}\sum_{\mathbf{k},m,l} V^{\mathrm{e-ph}}(\omega-\omega_m) \frac{i\omega_mZ(i\omega_m)}{\Theta_l(\mathbf{k},i\omega_m)} \nonumber  \\
&&-\frac{1}{2\omega}\int_{-\infty}^{\infty} \!\! \mathrm{d}z\sum_{\mathbf{k},l} \frac{\alpha^2F_l(z)}{N_l(0)} \frac{Z(\omega-z)(\omega-z)}{\Theta_l(\mathbf{k},\omega-z)} \zeta(\omega,z) \, ,\label{zomega}\\
&&\chi(\omega) = T\sum_{\mathbf{k},m,l} V^{\mathrm{e-ph}}(\omega-\omega_m) \frac{\xi_l(\mathbf{k})-\mu+\chi(i\omega_m)}{\Theta_l(\mathbf{k},i\omega_m)}\nonumber \\
&&+\int_{-\infty}^{\infty} \!\! \mathrm{d}z\sum_{\mathbf{k},l} \frac{\alpha^2F_l(z)}{2N_l(0)} \frac{\xi_l(\mathbf{k})-\mu+\phi(\omega-z)}{\Theta_l(\mathbf{k},\omega-z)} \zeta(\omega,z) ,~~~~\\
&&\phi(\omega) = -T\sum_{\mathbf{k},m,l} V^{\mathrm{e-ph}}(\omega-\omega_m) \frac{\phi(i\omega_m)}{\Theta_l(\mathbf{k},i\omega_m)} \nonumber \\
&& +\int_{-\infty}^{\infty} \mathrm{d}z\sum_{\mathbf{k},l} \frac{\alpha^2F_l(z)}{2N_l(0)} \frac{\phi(\omega-z)}{\Theta_l(\mathbf{k},\omega-z)} \zeta(\omega,z) ~, \label{phiomega}
\end{eqnarray}
with $\zeta(\omega,z)=\tanh \left(\frac{\omega-z}{2T} \right) +\coth \left(\frac{z}{2T} \right) $, introduced for brevity \cite{Marsiglio1988}. The solutions to Eqs.\,(\ref{zomega}-\ref{phiomega}) determine the real-frequency matrix Green's function of the system, which in turn can be used to find the momentum, band and frequency resolved spectral function:
\begin{eqnarray}
A_l(\mathbf{k},\omega)=-\frac{1}{\pi} \Im\left( \big[ \hat{G}_l(\mathbf{k},\omega+i\delta) \big]_{11} \right) \, . \label{spectral}
\end{eqnarray}
Here $\delta$ can be interpreted as a physical broadening. Eq.\,(\ref{spectral}) can in principle be directly compared to ARPES experiments. Further, the STS spectra are calculated by
\begin{eqnarray}
\frac{\mathrm{d}I}{\mathrm{d}V} \propto A(\omega)= \sum_{\mathbf{k},l} A_l(\mathbf{k},\omega) ~. \label{didv}
\end{eqnarray}
All selfconsistent equations presented here are solved using the Uppsala Superconductivity (UppSC) code \cite{UppSC,Aperis2015,Aperis2018}. Within the iterative loop we employ Fast Fourier transformation techniques for efficient summations and an Analytical Tail (AT) scheme for faster convergence in the number of Matsubara frequencies \cite{Schrodi2019}.

\section{Mapping between chemical potential and band filling}

\begin{figure}[h!]
\centering
    \renewcommand\thefigure{S1}
\includegraphics[width=0.9\linewidth]{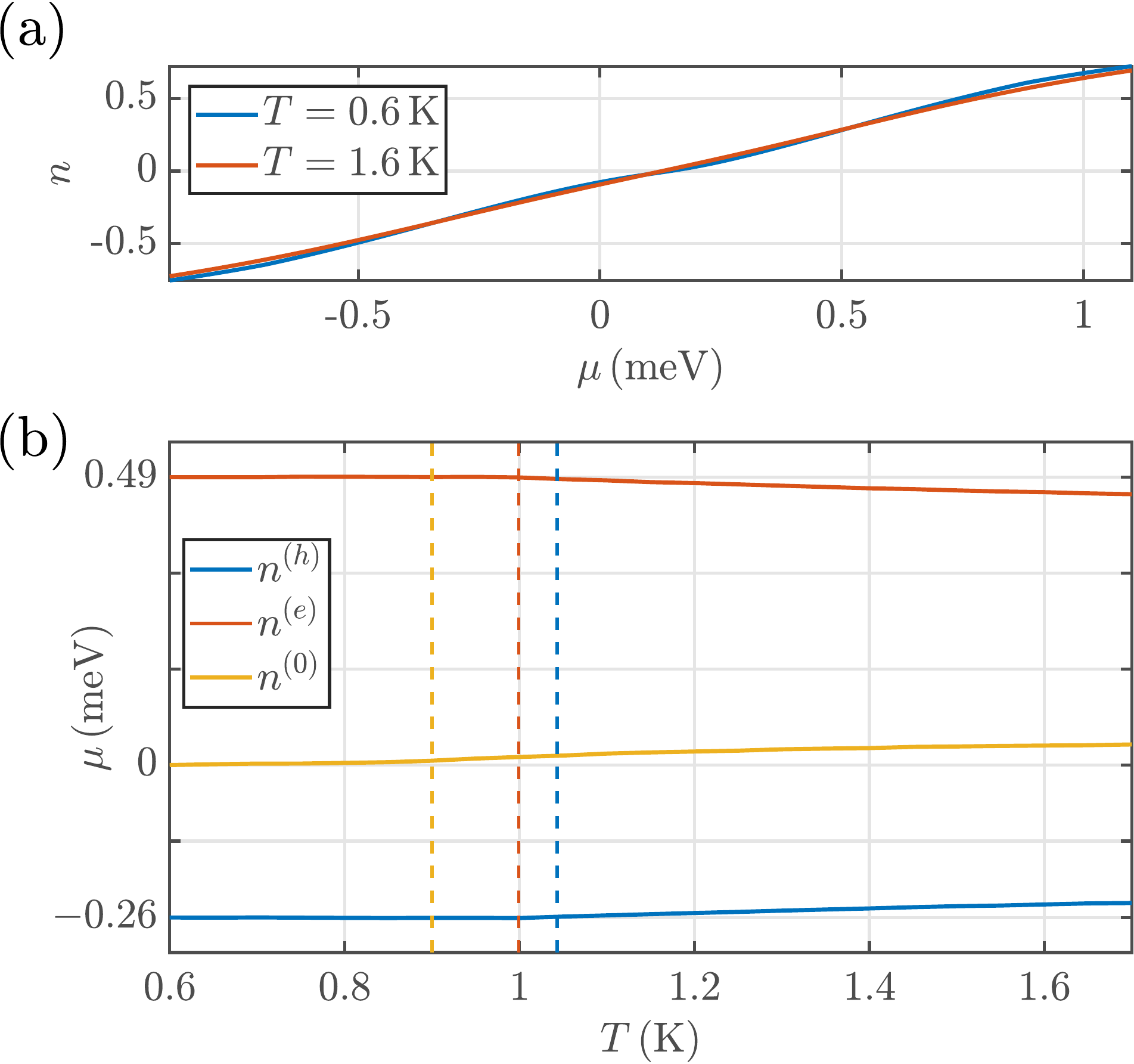}
	\caption{(a) Mapping between filling $n$ and chemical potential $\mu$ for $T=0.6\,\mathrm{K}<T_c$ in blue and $T=1.6\,\mathrm{K}>T_c$ in red. (b) Chemical potential evolution with temperature at three constant
		fillings $n^{(h)}$ (blue), $n^{(e)}$ (red) and $n^{(0)}$ (yellow). Dashed lines with the same color code correspond to respective transition temperatures.}	\label{mapping}
\end{figure}

We are interested in studying 
particular band fillings (especially $n^{(e)}$, $n^{(h)}$ and $n^{(0)}$ in the main text) that should remain constant with $T$.
Therefore, we allow a temperature dependent chemical potential for each $n$ {($n\in[-1,1]$).} Note, that we  introduced here a shift to have half-filling at $n=0$. Fig.\,\ref{mapping}(a) shows the relation between  $\mu$ and the electron filling $n$. The red curve is obtained in the normal state at $T=1.6\,\mathrm{K}>T_c$ and shows an approximately linear behavior. In the superconducting state (blue line, $T=0.6\,\mathrm{K}$) a slight distortion is observed, stemming from a finite value of the gap.

In panel (b) of Fig.\,\ref{mapping} we draw the temperature dependence of $\mu$ when keeping the filling constant. $n=n^{(e)}$, $n^{(h)}$ and $n^{(0)}$ are respectively shown by solid red, blue, and yellow curves. The three different transition temperatures are marked as dashed vertical lines in similar color code. The variation of $\mu$ with $T$ in each of the three cases is relatively minor. Interestingly, the chemical potentials are completely constant in the superconducting state. This behavior points towards small thermal renormalization effects on the band structure at $T>T_c$.
	
\section{Phase diagram including all electron bands}

\begin{figure}[t!]
	\centering
	    \renewcommand\thefigure{S2}
	\includegraphics[width=\linewidth]{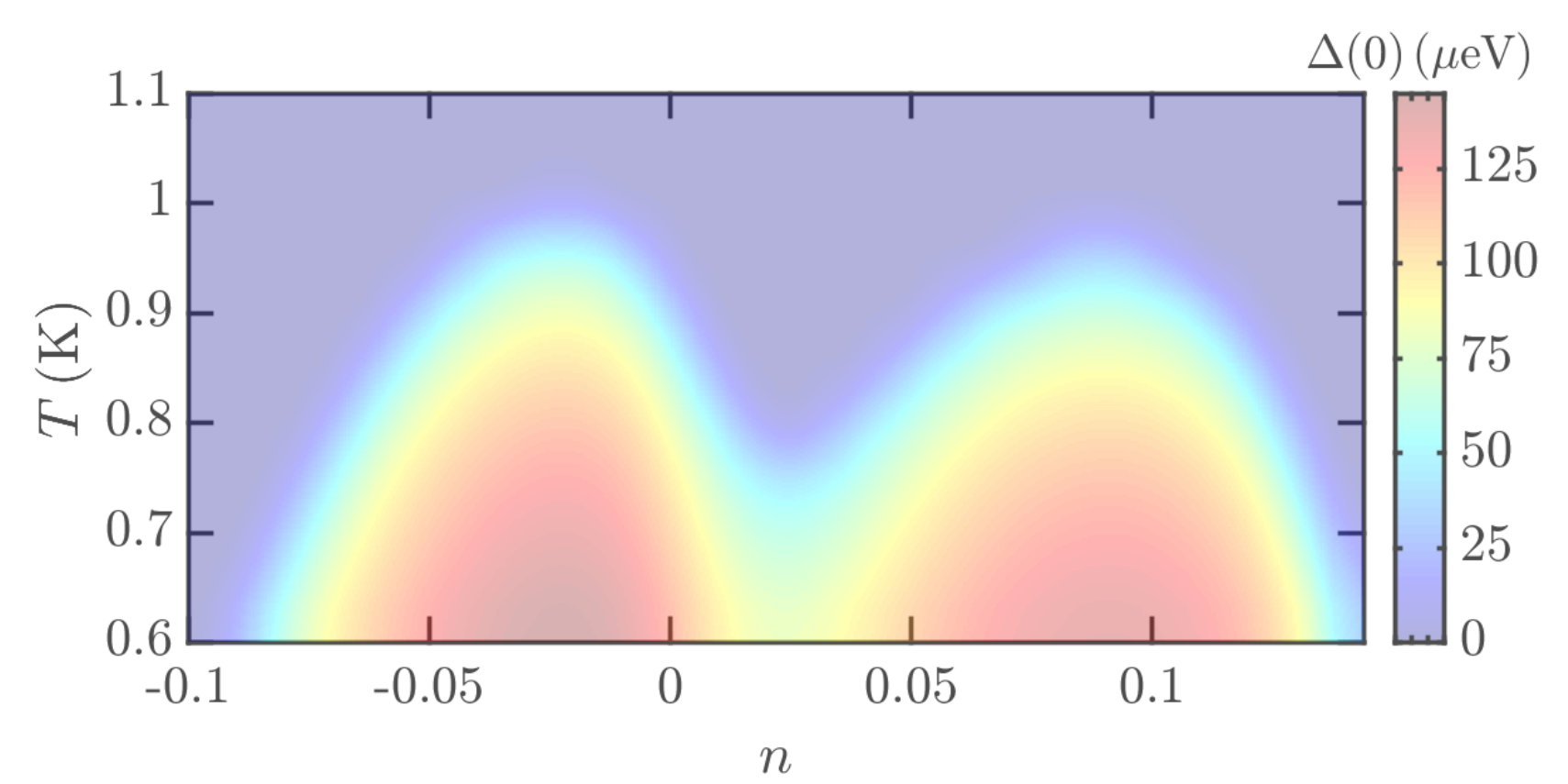}
	\caption{Phase diagram for $\Delta(0)$ as function of band filling and temperature, obtained from calculations with all 10 electronic bands.}\label{dome10}
\end{figure}

As described in the main text we employ a ten-band tight-binding description to model the electronic dispersion\,\cite{Po2019}, while in our main calculations we keep only the two flat bands close to the Fermi level. We have checked carefully that including the full range of energies does not alter our results significantly. To show this explicitly we repeated solving the Eliashberg equations as function of $n$ and $T$ by considering all 10 electronic energy bands. The resulting phase diagram is drawn in Fig.\,\ref{dome10}.
A comparison of this graph with Fig.\,2(a) in the main text reveals that the filling dependent values for $T_c$, as well as the maximum gap sizes essentially do not change. Note, that there is a difference on the axis for the electron filling, due to taking all bands into account.
Despite this aspect we obtain, as in the two-band case, two superconducting domes with particle-hole asymmetric gap sizes and critical temperatures. From this observation we are confident that the main physics is contained in the two flat bands close to the Fermi level. 

\section{Spectroscopy}

\begin{figure}[h]
	\centering
	    \renewcommand\thefigure{S3}
	\includegraphics[width=\linewidth]{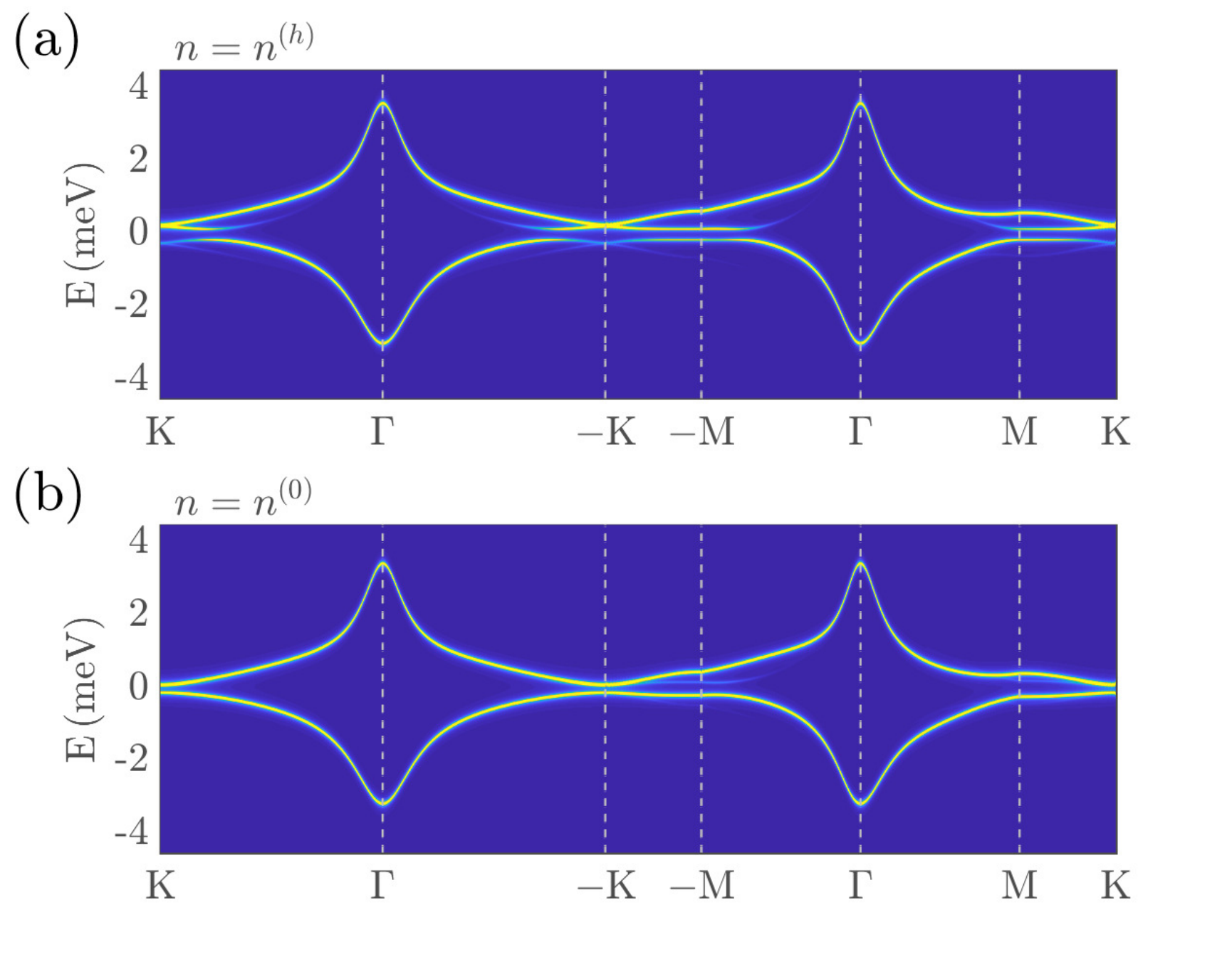}
	\caption{Spectral function along high-symmetry lines of the BZ, calculated at $T=0.2\,\mathrm{K}<T_c$. Electron fillings are chosen as $n^{(h)}$ in panel (a) and $n^{(0)}$ in (b).}	\label{arpes}
\end{figure}

Using Eq.\,(\ref{spectral}) we have access to the band, energy and momentum dependent spectral function $A_l(\mathbf{k},\omega)$, which can be probed by ARPES measurements. We show this quantity along high-symmetry lines in Figs.\,\ref{arpes}(a) and (b) for electron fillings $n^{(h)}$ and $n^{(0)}$, respectively.

\begin{figure}[h!]
	\centering
	    \renewcommand\thefigure{S4}
	\includegraphics[width=\linewidth]{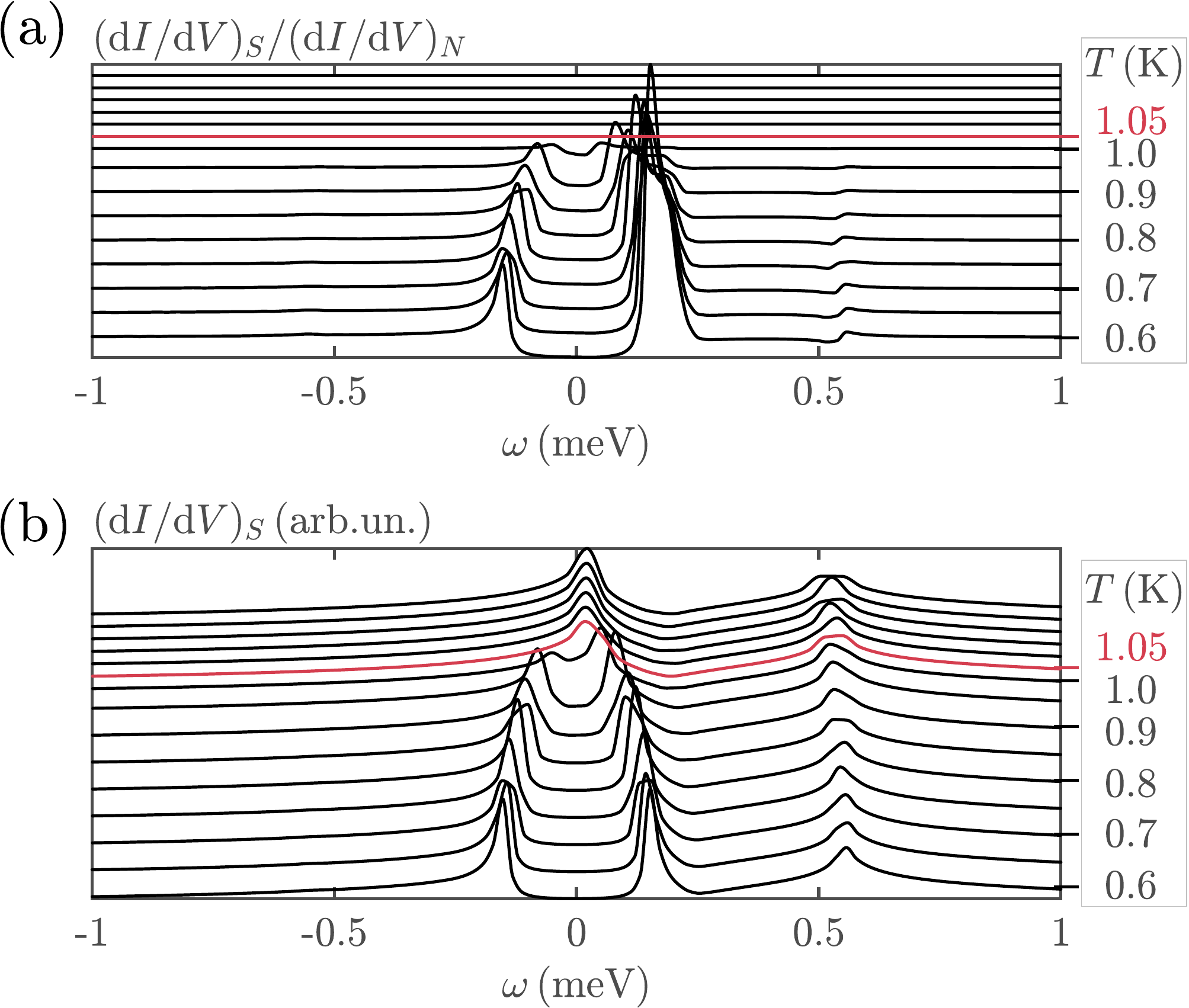}
	\caption{Temperature dependent differential conductance at electron filling $n=n^{(h)}$. (a) Normalized and (b) non-normalized spectra.}	\label{tunneling}
\end{figure}

An experimentally observable tunneling spectrum can be calculated from Eq.\,(\ref{didv}). This is done by normalizing $(\mathrm{d}I/\mathrm{d}V)_S$ in the superconducting state by $(\mathrm{d}I/\mathrm{d}V)_N$ of the normal state ($T>T_c$). In Fig.\,\ref{tunneling} we show the calculated temperature dependent outcome for filling $n^{(h)}$. Panel (a) and (b) show, respectively, the normalized and nonnormalized differential conductance.
\\

\bibliographystyle{apsrev4-1}
%

\end{document}